\documentclass[prl,aps,twocolumn,showpacs]{revtex4}
\usepackage{epsfig}
\include{graphics}
\begin{document}
\title{Scaling anomalies in the coarsening dynamics of fractal viscous fingering patterns}
\author{Massimo Conti$^{1}$, Azi Lipshtat$^{2}$ and Baruch Meerson$^{2}$}
\affiliation{$^{1}$Dipartimento di Matematica e Fisica,
Universit\'{a} di Camerino, and Istituto Nazionale di Fisica della
Materia, 62032, Camerino, Italy} \affiliation{$^{2}$Racah
Institute  of  Physics, Hebrew University of Jerusalem, Jerusalem
91904, Israel}
\begin{abstract}
We analyze a recent experiment of Sharon \textit{et al.} (2003) on
the coarsening, due to surface tension, of fractal viscous
fingering patterns (FVFPs) grown in a radial Hele-Shaw cell. We
argue that an unforced Hele-Shaw model, a natural model for that
experiment, belongs to the same universality class as model B of
phase ordering. Two series of numerical simulations with model B
are performed, with the FVFPs grown in the experiment, and with
Diffusion Limited Aggregates, as the initial conditions. We
observed Lifshitz-Slyozov scaling $t^{1/3}$ at intermediate
distances and very slow convergence to this scaling at small
distances. Dynamic scale invariance breaks down at large
distances.

\end{abstract}
 \maketitle

Coarsening is an important paradigm of emergence of order from
disorder. It has been extensively studied in two-phase systems
quenched from a disordered state into a region of phase
coexistence \cite{Lifshitz,LS,Gunton}.  
In another class of systems, disordered configurations are
generated by an instability of growth in combination with noise,
and they often exhibit long-range correlations and fractal
geometry \cite{fractals}. Examples include fractal clusters
developing in the process of solidification from an under-cooled
liquid \cite{Langer}, fractal clusters on a substrate grown by
deposition \cite{Landman} and fractal viscous fingering patterns
(FVFPs) formed by the Saffman-Taylor instability in the radial
Hele-Shaw cell \cite{Langer}. When the driving stops, the fractal
clusters coarsen by surface tension, and the coarsening dynamics
provide a valuable characterization of these systems.

An important simplifying factor in the analysis of coarsening
dynamics is \textit{dynamic scale invariance} (DSI): the presence
of a \textit{single}, time-dependent length scale $L(t)$, so that
a normalized pair correlation function $C(r,t)$ depends, at long
times, only on $r/L(t)$. The coarsening length scale $L(t)$ often
exhibits a power law in time \cite{Gunton}. For systems with
short-range correlations there is a lot of evidence, from
experiments and numerical simulations, in favor of DSI
\cite{Gunton}. For systems with long-range correlations the
situation is more complicated. In the case of a non-conserved
order parameter DSI was established in particle simulations
following a quench from $T=T_{c}$ to $T=0$ \cite{Humayun}.
Implications of mass conservation in DSI were addressed more
recently, in the context of coarsening of fractal clusters. Most
remarkable of them is the predicted decrease of the cluster radius
with time \cite{Sempere}. As of present, only the systems where
the conservation law is imposed \textit{globally}, rather than
locally, have been found to indeed show this effect \cite{PCM}. On
the contrary, the ``frozen" structure of fractal clusters at large
distances, observed in simulations of \textit{locally} conserved
(diffusion-controlled) fractal coarsening
\cite{Irisawa,CMS,Kalinin,LMS} implies breakdown of DSI in these
systems \cite{CMS,LMS}. The frozen structure is due to Laplacian
screening of transport at large distances \cite{LMS}.

An additional scaling anomaly, observed in the numerical
simulations of diffusion-controlled fractal coarsening
\cite{Irisawa,CMS,Kalinin,LMS}, was the presence of \textit{two}
apparently different dynamic length scales. For one of them,
determined from the time-dependence of either the slope of the
Porod-law part of $C(r)$ \cite{CMS,Kalinin,LMS}, or the cluster
perimeter \cite{Irisawa,CMS,LMS}, a power law in time was
reported: $L_1 \sim t^{0.20 - 0.23}$. Another length scale,
determined from a knee-like structure in $C(r)$ at moderate
distances  behaves like $L_2 \sim t^{0.30 - 0.32}$ \cite{LMS}.
While $L_2(t)$ can be identified as Lifshitz-Slyozov length scale
$\sim t^{1/3}$ \cite{LMS}, the length scale $L_1(t)$ looks
unusual.

Strikingly similar results were recently obtained in experiment on
the coarsening dynamics of a \textit{different} system: radially
grown FVFPs in a Hele-Shaw cell \cite{Sharon}. The frozen
structure at large distances, observed in Ref. \cite{Sharon},
clearly indicates breakdown of DSI. Furthermore, two different
time-dependent length scales, with apparent dynamic exponents
$0.22$ and $0.31$, are observed \cite{Sharon}. Why is this system
so similar to the diffusion-controlled system? Where does the
exponent $0.20-0.23$ come from? These questions are addressed in
the present work. We first suggest an unforced Hele-Shaw model and
discuss its properties. A scaling argument indicates that this
model belongs to the same universality class as the so called
model B, the standard model of the diffusion-controlled phase
separation in two-phase systems \cite{Gunton}. Assuming
universality, we performed a series of numerical simulations with
model B, where the FVFPs, grown in experiment of Sharon \textit{et
al.} \cite{Sharon}, are used as the initial conditions for the
minority phase. Then we report additional simulations of fractal
coarsening, with DLAs (Diffusion Limited Aggregates) as the
initial conditions. These two series of simulations show
Lifshitz-Slyozov scaling $t^{1/3}$ at intermediate distances.
Breakdown of dynamic scale invariance at large distances is
confirmed. However, the existence of an anomalous power law in
$L_1(t)$ is disproved.

A natural description of coarsening of FVFPs is provided in terms
of an unforced Hele-Shaw (UHS) flow. Consider a Hele-Shaw flow
\cite{Saffman} and assume that the driving fluid (for example,
air) has negligible viscosity, so that the pressure inside it is
spatially uniform. When the plate spacing $b$ is very small, the
flow is effectively two-dimensional, and the velocity of the
viscous fluid (for example, oil) is
$\mathbf{v}\,(\mathbf{r},t)=-(b^2/12\mu) \,\nabla
p\,(\mathbf{r},t)$, where $p$ is the pressure and $\mu$ is the
dynamic viscosity of the driven fluid. As the fluids are
immiscible, the interface speed is
\begin{equation}\label{speed}
  v_n = - \frac{b^2}{12 \mu} \nabla_n p\,,
\end{equation}
where index $n$ denotes the components of the vectors normal to
the interface, and the gradient is evaluated at the respective
points of the interface $\gamma$. Assuming incompressibility of
the driven fluid, $\mathbf{\nabla} \cdot \,\mathbf{v}=0$, one
arrives at Laplace's equation for the pressure:
\begin{equation}\label{Laplace}
  \nabla^2 p =0\,.
\end{equation}
The pressure jump across the interface is \cite{Park}
\begin{equation}\label{jump1}
  \Delta p  = \frac{\sigma}{b} \left[1+3.8 \left( \frac{\mu
  v_n}{\sigma}\right)^{2/3}\right]+\frac{\pi}{4}\,\sigma {\cal K}\,,
\end{equation}
where $\sigma$ is surface tension, and ${\cal K}$ is the curvature
of the interface. At the coarsening stage the interface speed is
very small, so the second term in the square brackets can be
neglected. The first term does not depend on the coordinates, so
one arrives at a Gibbs-Thomson relation
\begin{equation}\label{jump2}
  \Delta p  = \frac{\pi}{4}\,\sigma {\cal K}\,.
\end{equation}
To close this set of equations, one more condition is needed.  A
natural condition to demand during the \textit{growth} stage is a
constant-in-time driving pressure \cite{Kadanoff}, or a constant
areal flow rate of the driving fluid. Each of these conditions
assumes evacuation of the driven fluid at the external boundary of
the system. In the coarsening problem both the supply of the
driving fluid, and evacuation of the driven fluid are blocked.
Therefore, the normal component of the velocity of the driven
fluid at the external boundary $\Gamma$ should vanish, which
follows
\begin{equation}\label{external}
\nabla_n p\,|_{\Gamma} = 0\,.
\end{equation}
Equations (\ref{speed}), (\ref{Laplace}), (\ref{jump2}) and
(\ref{external}) define a one-sided version of the UHS model.
Similar models have been used in the context of break-ups
(pinch-offs) of bubbles, driven only by surface
tension~\cite{Almgren,afterbreakup}. The UHS model has two
important properties: (i) The total area $A$ of the
driving fluid is constant. 
(ii) The total length of the interface is a non-increasing
function of time~\cite{Constantin}.

Now let us compare the UHS model with model B, the phase-field
formulation of which is given by the Cahn-Hilliard equation for
the order parameter $u(\mathbf{r},t)$~\cite{Gunton}:
\begin{equation}
\frac{\partial u}{\partial t} + \frac{1}{2}\nabla^2 \left(\nabla^2 u + u - u^3 \right) = 0 \,.
\label{CH}
\end{equation}
At late times, the two-phase dynamics are describable by an
asymptotic sharp-interface theory \cite{Pego}. In the
sharp-interface limit, the interface speed is
\begin{equation}\label{speedCH}
v_n =  \frac{1}{4} \left(- \nabla_n \Phi^{out}+\nabla_n
\Phi^{in}\right)\,,
\end{equation}
where potential $\Phi(\mathbf{r},t)$ is a harmonic function in
each of the two phases \textit{in} and \textit{out}. The boundary
conditions are $\Phi|_{\gamma}  = (\sqrt{2}/3)\,{\cal K}$ and
$\nabla_n \Phi\,|_{\Gamma} = 0$.

How are these two problems related? To begin with, the
sharp-interface limit of model B has the same properties (i) and
(ii) as the UHS model \cite{Pego}, so each of the two models
describes interface-shortening dynamics under area conservation.
The models do differ from each other considerably in the
\textit{final} outcomes of the coarsening dynamics. For a steady
state solution of the UHS model one has simply $p=const$.
Therefore, possible stationary shapes of domains of the driving
fluid in the UHS model are \textit{one or more} circular bubbles
of arbitrary radii. On the contrary, in model B, $\Phi=const$
cannot be a steady state solution in the presence of more than one
bubbles, because it cannot obey all the boundary conditions on the
multiple interfaces. Therefore, a generic final state here is
always a \textit{single} circular bubble. In model B bubbles
compete for material via diffusion through the majority phase.
Obviously, this competition mechanism (Ostwald
ripening~\cite{LS,Gunton}) is absent in the UHS model.

This difference between the two models becomes crucial after the
driving fluid breaks up into multiple bubbles. Before it happens,
the two models can be expected to behave similarly. A simple
argument for this follows from scaling analysis. Consider
coarsening of a domain of complex shape and assume for a moment
DSI, that is a \textit{single} relevant length scale $L=L(t)$. The
pressure jump across the interface can be estimated from Eq.
(\ref{jump2}): $\Delta p \sim \sigma/L$. Then, from  Eq.
(\ref{speed}), $v_n \sim b^2 \sigma/(\mu L^2)$. On the other hand,
$v_n \sim \dot{L}$. This yields a coarsening law $L(t) \sim (b^2
\sigma t/\mu)^{1/3}$. We checked that this estimate is in
excellent agreement with the experimental result~\cite{Sharon} for
$L_2(t)$, for two latest decades of time.

The same power law $t^{1/3}$ is obtained in model B
\cite{LS,Gunton}. Therefore, if DSI holds, the two models belong
to the same universality class. In reality, each of these two
systems exhibits breakdown of DSI at large distances, when one
deals with fractal clusters at
$t=0$~\cite{CMS,Kalinin,LMS,Sharon}. At intermediate distances,
however, the classic exponent $1/3$ \textit{is} observed in both
systems ~\cite{LMS,Sharon}. Therefore, we conjecture that, prior
to major breakup, the two models belong to the same universality
class.

Based on this conjecture, we performed two series of simulations
with model B [Eq. (\ref{CH})]. Details of our numerical procedure
and diagnostics can be found in Refs.~\cite{CMS,LMS}. In the first
series of simulations we used FVFPs grown in experiment
\cite{Sharon} as the initial conditions for the ``minority phase"
$u=1$. The fractal dimension of these patterns, determined from
the pair correlation function, is close to $1.71$. The (scaled)
system size was $1024 \times 1024$, with periodic boundary
conditions. The (scaled) time range of the simulations was $0<t<3
\cdot 10^4$. Figure 1 shows snapshots of the simulated coarsening
dynamics. The snapshots closely resemble those observed in
experiment~\cite{Sharon}. Figure 2 presents the (normalized)
equal-time pair correlation function $C(r,t)$ at different times,
and the characteristic dynamic length scales. The data is averaged
over 7 simulations with different FVFPs. $C(r,t)$ in a linear
scale is shown in Fig. 2a. At small distances $C$ goes down
linearly with $r$ (the Porod law) \cite{Gunton}, and the inverse
slope of this linear dependence yields the ``coarsening length
scale" $L_1(t)$ depicted in Fig. 2d. The log-log plots of $C(r,t)$
(Fig. 2b) indicate an invariable fractal dimension of the cluster
at large distances (up to the upper cutoff of the fractal). In
addition, Fig. 2b exhibits a knee-like feature. In the previous
work \cite{LMS} a similar knee-like feature served to identify
Lifshitz-Slyozov length scale $L_2(t)$. Here, following Sharon
\textit{et al}.~\cite{Sharon}, we subtracted from $C(r,t)$ its
initial value $C(r,0)$, and followed the dynamics of the
difference, see Fig. 2c. The knee-like feature of Fig. 2b becomes
here a local minimum whose position at different times yield a
sharp estimate of $L_2(t)$. The ``frozen" tail at the distances
much larger than $L_2(t)$, but still much smaller than the system
size, implies breakdown of DSI.

\begin{figure}[ht]
\begin{tabular}{cc}
  \epsfxsize=3.5cm  \epsffile{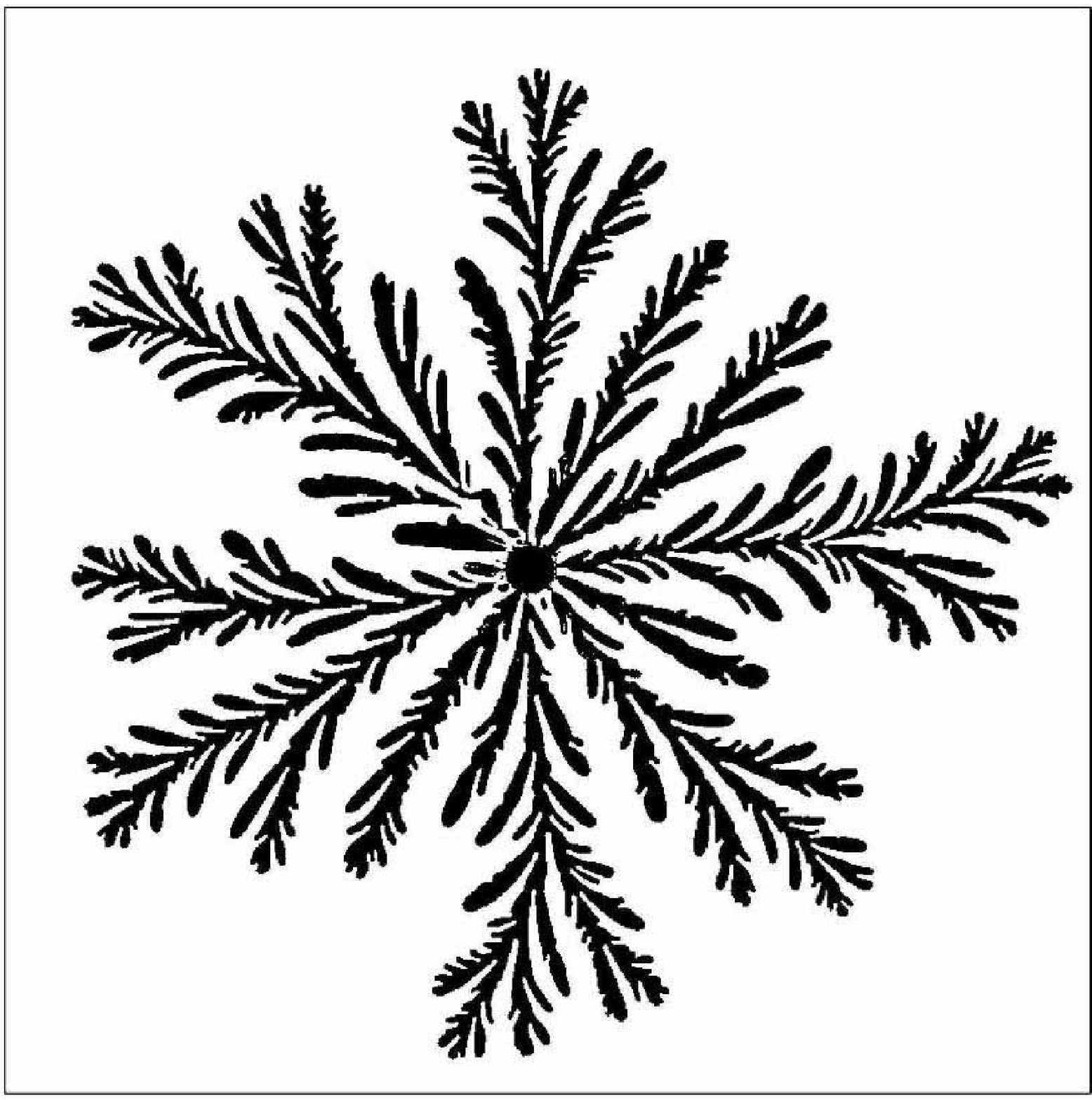} & \epsfxsize=3.5cm  \epsffile{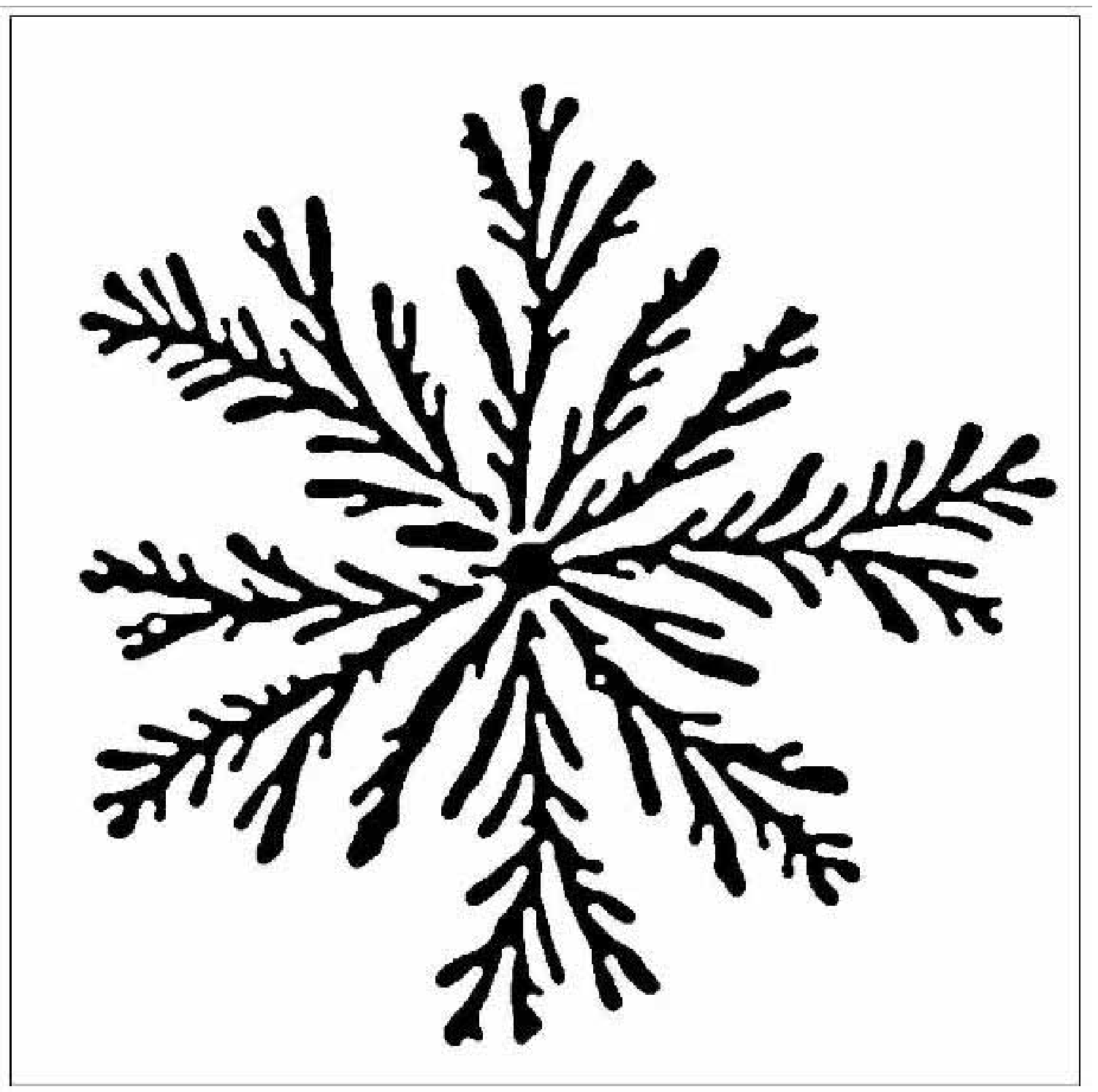}\\
  \epsfxsize=3.5cm  \epsffile{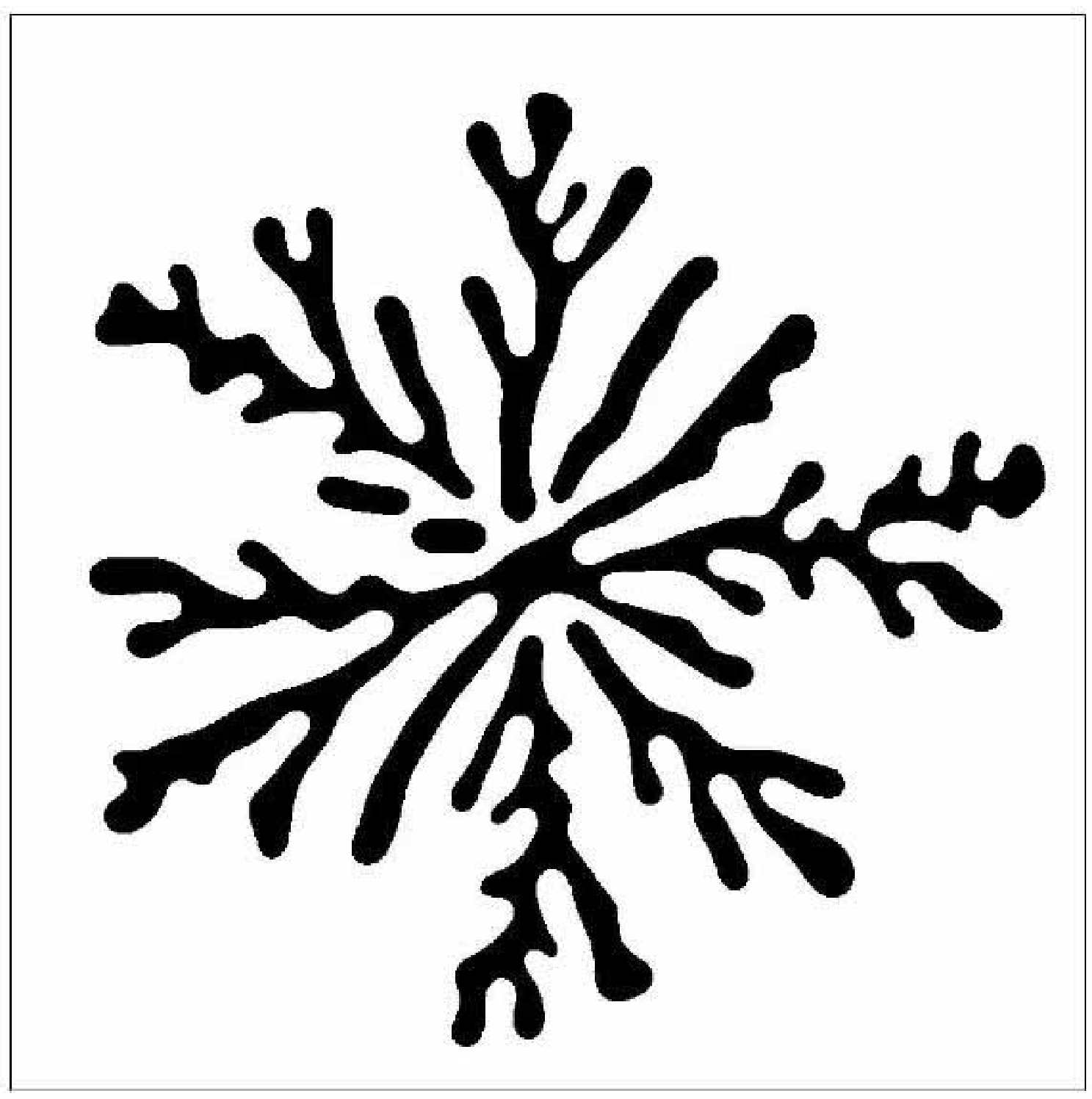} & \epsfxsize=3.5cm  \epsffile{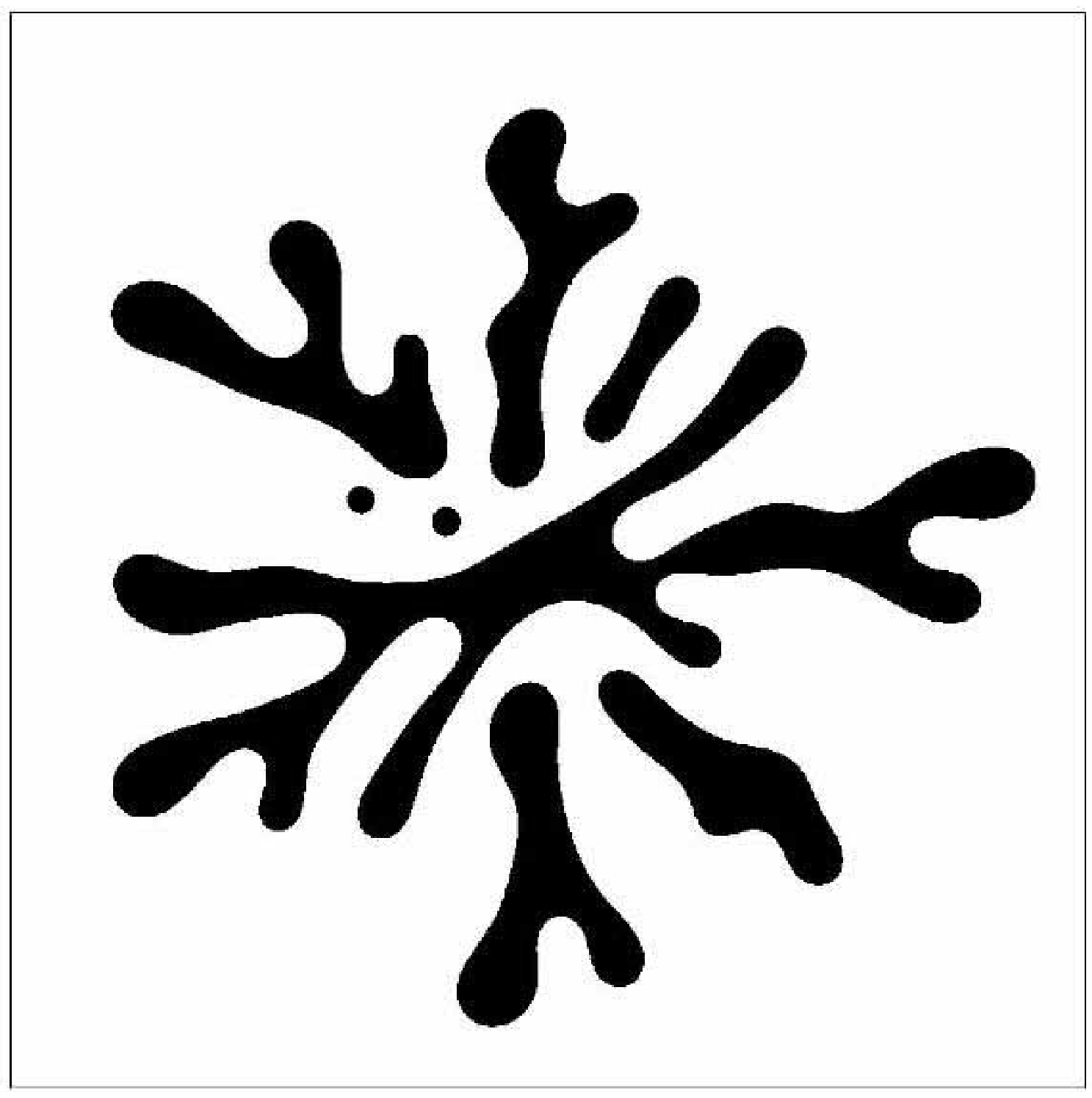}\\
\end{tabular}
\caption{Snapshots of coarsening of FVFPs simulated with model B.
The upper left figure ($t=0$) shows a FVFP ($D \simeq 1.71$) grown
in experiment~\cite{Sharon}. The rest of the snapshots show the
simulation results at scaled times $t=290$ (upper right), $3817$
(lower left) and $30 000$ (lower right).} \label{fig1}
\end{figure}

\begin{figure}[ht]
\epsfig{file=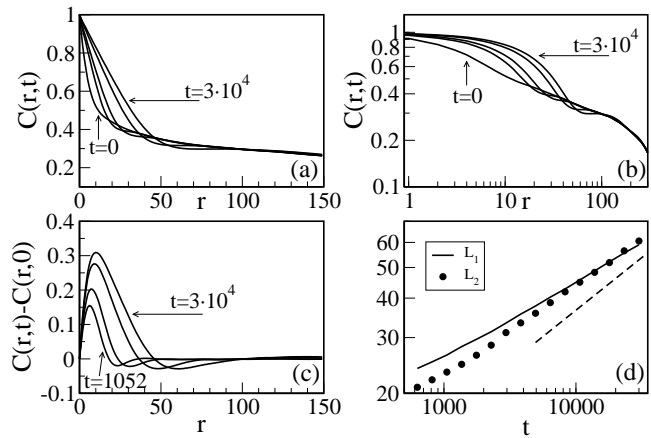,angle=270,width=8.5cm, clip=} \caption{The
dynamics of the equal-time pair correlation function $C(r,t)$
(a-c), and the dynamic length scales $L_1(t)$ and $L_2(t)$ (d)
from simulations with model B. The initial conditions are the
FVFPs grown in experiment~\cite{Sharon}. The time moments are
$t=0$ (figures a and b only), $1052$, $2950$, $13 846$ and $3
\cdot 10^4$. The dashed line describes a power law $\sim t^{1/3}$
and is shown here to guide the eye. See text for further details.}
\label{fig2}
\end{figure}

Power-law fits of the data shown in Fig 2d yield the following
dynamic exponents: $\alpha_1=0.24 \pm 0.01$ for $L_1 \sim
t^{\alpha_1}$ and $\alpha_2=0.30 \pm 0.01$ for $L_2 \sim
t^{\alpha_2}$. The same result for $\alpha_1$ is obtained from a
power-law fit of the cluster perimeter versus time $P(t) \sim
t^{-\alpha_1}$, as expected \cite{Gunton,Irisawa,CMS,PCM,LMS}.
While the value of $\alpha_2$ is very close to that obtained in
earlier simulations of model B ~\cite{LMS} and experiment with the
FVFPs~\cite{Sharon}, $\alpha_1=0.24$ is somewhat larger than the
values  $0.20-0.23$ reported earlier for these two
systems~\cite{Irisawa,CMS,Kalinin,LMS,Sharon}. Also, noticeable in
Fig. 4d is curvature of the log-log plot of $L_1(t)$.  These
observations put forward a question about the true asymptotic
value of exponent $\alpha_1$.

To address this question, we performed a series of larger
simulations, extending the time interval until $t=10^5$ (which is
$20$ times longer than the first phase-field simulations of this
system \cite{CMS}). The initial conditions for the minority phase
were DLA clusters, ``reinforced" by an addition of peripheral
sites. The clusters occupied the $1024\times1024$ box; they had a
larger fractal range than the FVFPs grown in
experiment~\cite{Sharon}. Figure 3 shows snapshots of the
simulated coarsening dynamics. Figure 4a presents $C(r,t)$
averaged over 6 different realizations of DLA. Again, following
some initial ``evaporation" of the minority phase (which happens
at an earlier stage of the Cahn-Hilliard dynamics), the tail of
$C(r)$ is frozen until very long times. The dynamic length scale
$L_2(t)$ is shown in Fig. 4b; a power law fit at long times yields
$\alpha=0.31 - 0.32$ which is close to $1/3$, as expected. The
cluster perimeter versus time is shown in Fig. 4c. It can be seen
that $P(t)$ has not approached yet a power law. Therefore, we
followed Huse~\cite{Huse} and introduced an effective
\textit{time-dependent} exponent $- \alpha_1 (t)$ which is shown
in Fig. 4d versus the perimeter $P$ itself. An asymptotic value of
$- \alpha_1(t)$ is obtained by linear extrapolation $t\to \infty$,
that is $P \to 0$. This procedure yields $\alpha_1 = 0.34$, very
close to $1/3$.

\begin{figure}[ht]
\begin{tabular}{cc}
  \epsfxsize=3.5cm  \epsffile{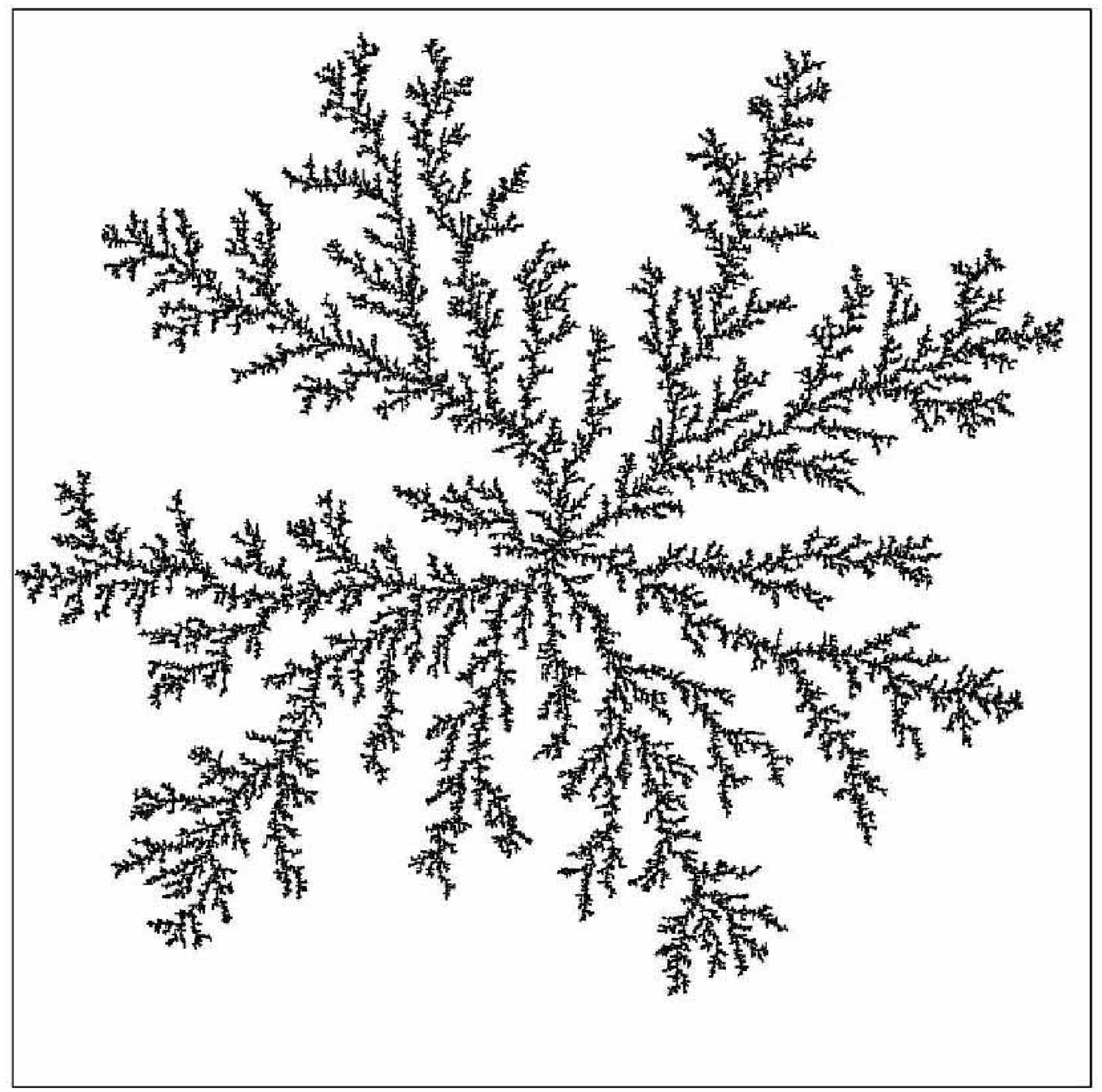} & \epsfxsize=3.5cm  \epsffile{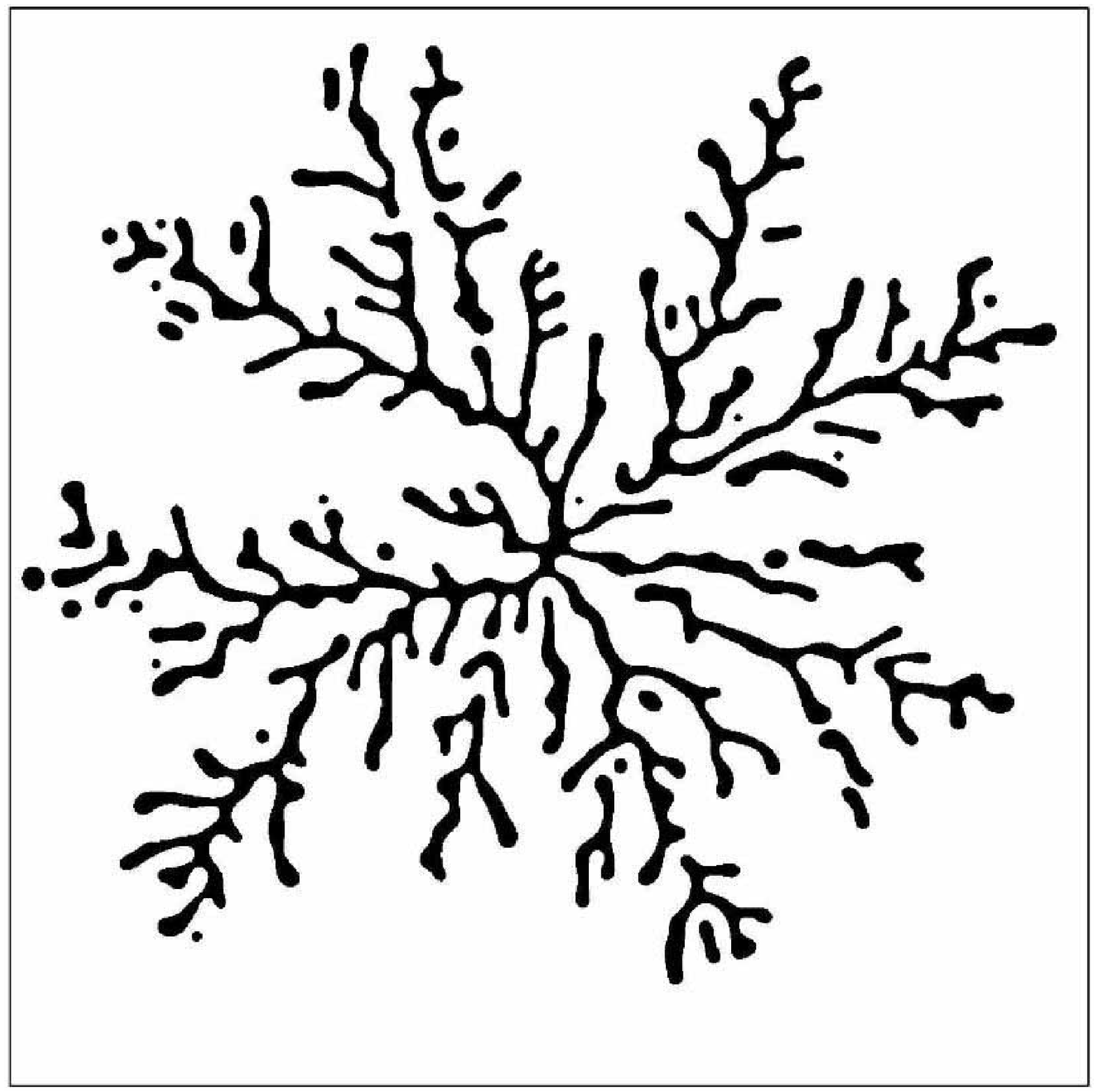}\\
  \epsfxsize=3.5cm  \epsffile{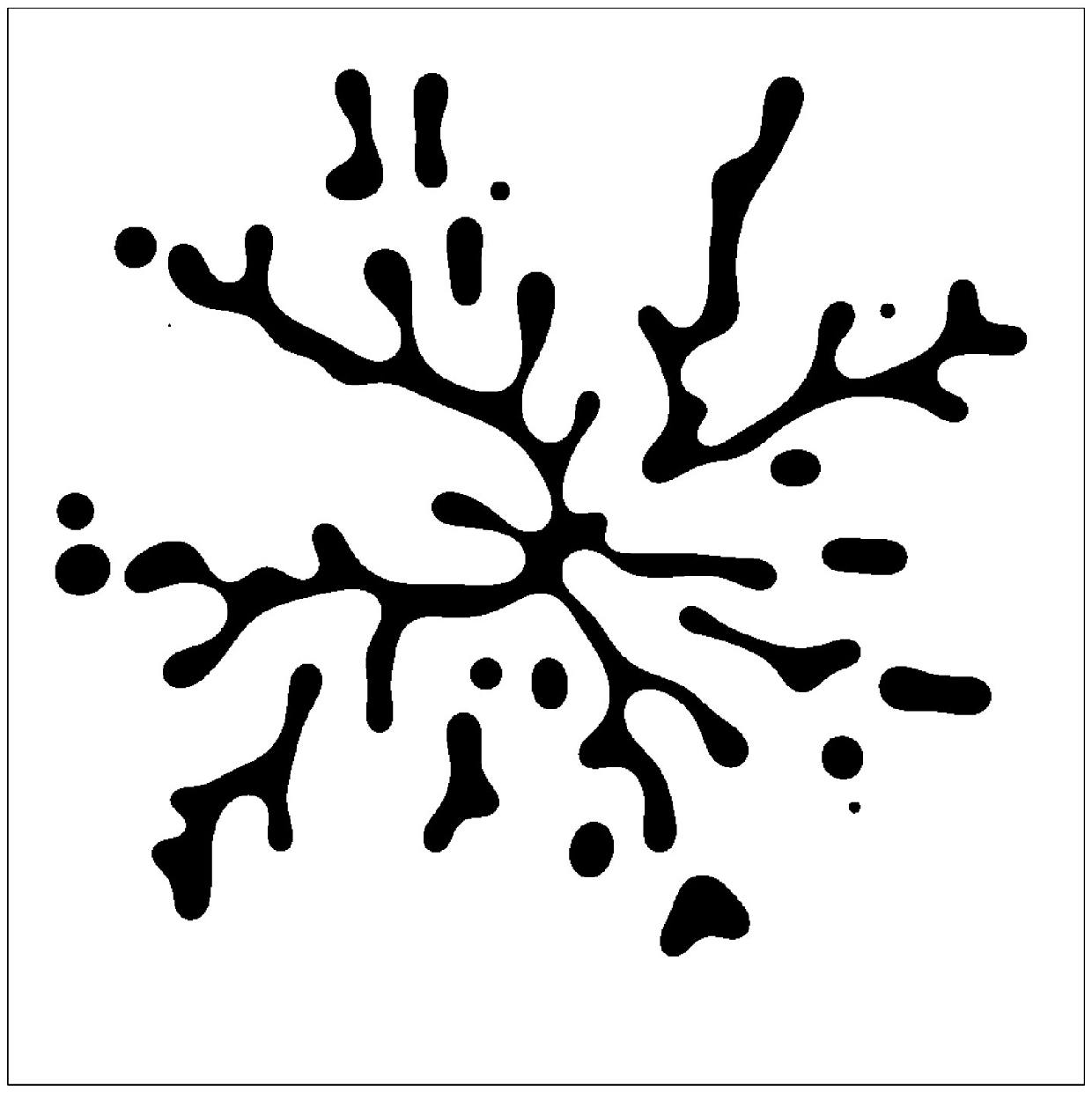} & \epsfxsize=3.5cm  \epsffile{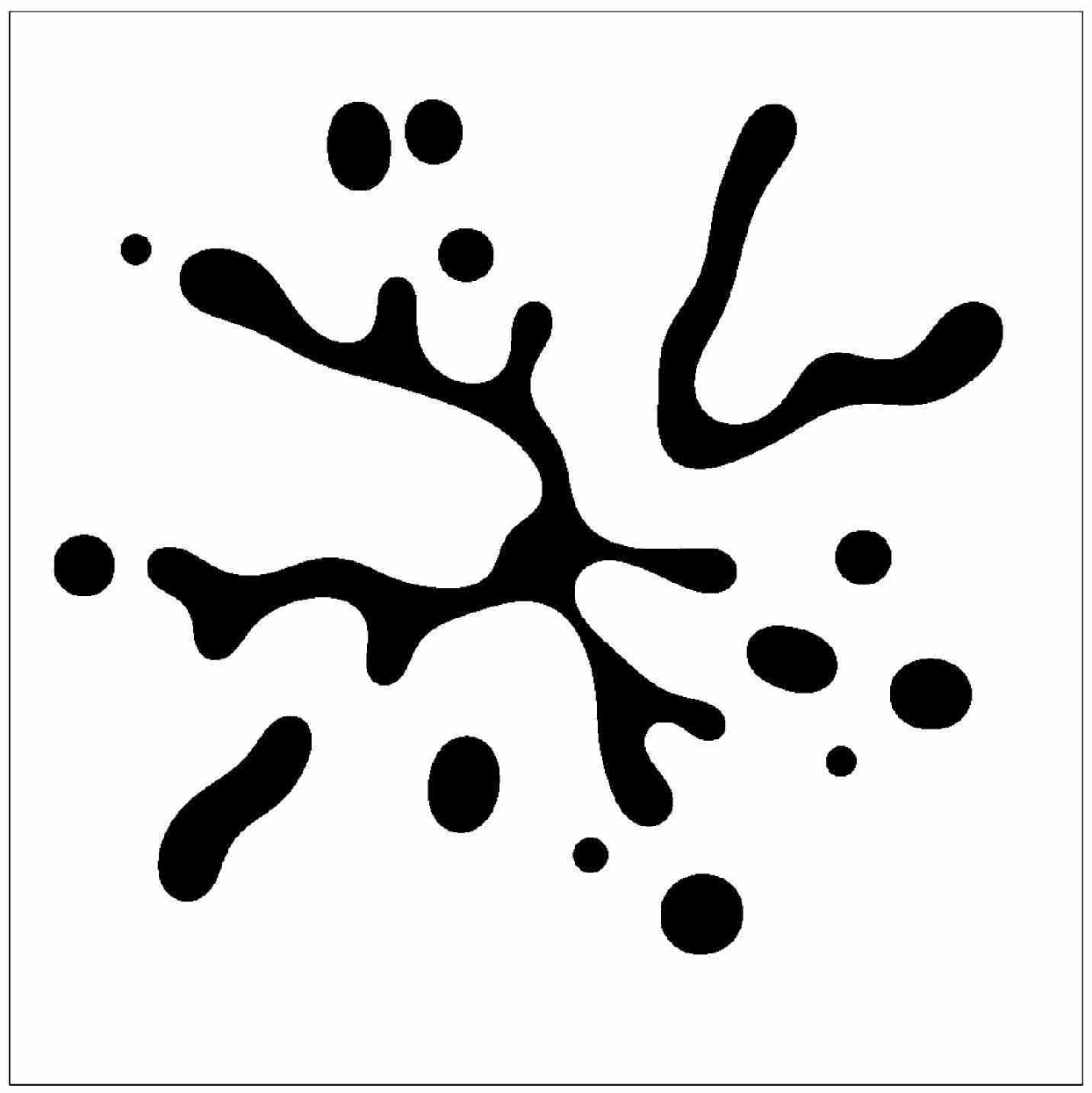}\\
\end{tabular}
\caption{Snapshots of coarsening of DLA clusters ($D\simeq 1.71$),
simulated with model B, at scaled times $t=0$ (upper left), $1350$
(upper right), $26 591$ (lower left) and $10^5$ (lower right).}
\label{fig3}
\end{figure}

\begin{figure}[ht]
\epsfig{file=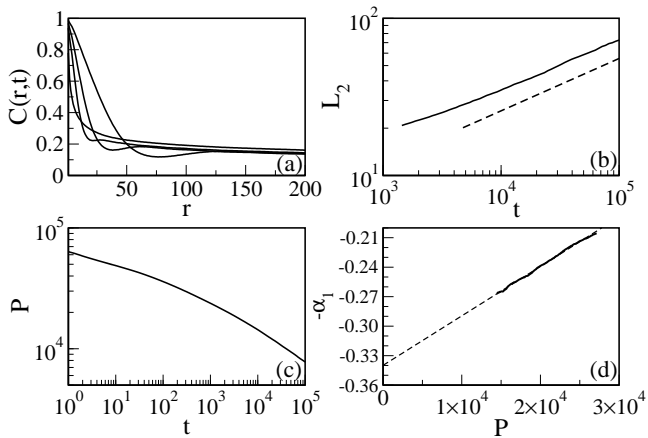,angle=270,width=8.5cm, clip=}
\caption{$C(r,t)$ at  $t=0$, $1026$, $10 521$ and $10^5$ (a),
$L_2(t)$ (b) and $P(t)$ (c). The dashed line describes a power law
$\sim t^{1/3}$ and is shown here to guide the eye. The effective
time-dependent exponent $\alpha_1$ versus $P$ is shown in Fig. d.
Linear extrapolation to $P = 0$ (dashed line) yields $\alpha_1 =
0.34$.} \label{fig4}
\end{figure}

Overall, our simulations with model B are in remarkable agreement
with experiment on the Hele-Shaw coarsening of FVFPs
\cite{Sharon}. Breakdown of DSI at large distances and the
Lifshitz-Slyozov scaling $t^{1/3}$ at intermediate distances are
firmly established. Our simulations show, however, that the
``unusual" dynamic exponent $0.20 - 0.23$ is a transient on the
way to $1/3$. This finding explains the apparent independence of
the unusual exponent on the fractal dimension of the cluster,
observed in Ref. \cite{LMS}. In view of the conjectured
universality, we expect that the same kind of behavior will be
observed in a larger-scale experiment on the coarsening of FVFPs,
and in direct simulations with the unforced Hele-Shaw model.

We are grateful to Eran Sharon and his colleagues for providing
images of FVFPs, that were used in our simulations, and for useful
discussions. We thank Avner Peleg and Boris Zaltzman for advice.
The work was supported by the Israel Science Foundation (grant No.
180/02).

\end{document}